\documentclass[prx,aps,twocolumn,superscriptaddress]{revtex4-2}

\usepackage[pdftex]{graphicx}
\usepackage{amsbsy,amssymb,amsmath,bm,mathtools}
\usepackage{amsfonts}
\usepackage{mathrsfs}
\usepackage{makecell}
\usepackage{physics}

\usepackage{xspace}
\usepackage{bm}
\usepackage{color}
\usepackage{url}
\usepackage[section]{placeins}
\usepackage[colorlinks=true,linkcolor=blue,citecolor=blue]{hyperref}

\usepackage{tikz}

\newcommand{\recttwobyone}{
\begin{tikzpicture}[scale=0.2,baseline=-0.5ex]
\draw (0,0) rectangle (2,1);
\draw (1,0) -- (1,1);
\end{tikzpicture}
}

\newcommand{\rectonebytwo}{
\mathrel{
\begin{tikzpicture}[scale=0.2,baseline=-0.6ex]
\draw (0,0) rectangle (1,2);
\draw (0,1) -- (1,1);
\end{tikzpicture}
}
}

\newcommand{\Lshape}{
\mathrel{
\begin{tikzpicture}[scale=0.2,baseline=-0.6ex]

\draw (0,0) -- (2,0);     
\draw (0,0) -- (0,2);     
\draw (0,2) -- (1,2);     
\draw (2,0) -- (2,1);     

\draw (1,0) -- (1,1);     
\draw (0,1) -- (1,1);     

\draw (1,1) -- (2,1);     
\draw (1,1) -- (1,2);     

\end{tikzpicture}
}
}

\begin{document} 

\title{Graph Neural Networks in the Wilson Loop Representation \\of Abelian Lattice Gauge Theories}

\author{Ali Rayat}
\affiliation{Department of Physics, University of Virginia, Charlottesville, VA 22904, USA}

\author{Gia-Wei Chern}
\affiliation{Department of Physics, University of Virginia, Charlottesville, VA 22904, USA}

\date{\today}

\begin{abstract}
Local gauge structures play a central role in a wide range of condensed matter systems and synthetic quantum platforms, where they emerge as effective descriptions of strongly correlated phases and engineered dynamics. We introduce a gauge-invariant graph neural network (GNN) architecture for Abelian lattice gauge models, in which symmetry is enforced explicitly through local gauge-invariant inputs, such as Wilson loops, and preserved throughout message passing, eliminating redundant gauge degrees of freedom while retaining expressive power. We benchmark the approach on both $\mathbb{Z}_2$ and $\mathrm{U}(1)$ lattice gauge models, achieving accurate predictions of global observables and spatially resolved quantities despite the nonlocal correlations induced by gauge–matter coupling. We further demonstrate that the learned model serves as an efficient surrogate for semiclassical dynamics in $\mathrm{U}(1)$ quantum link models, enabling stable and scalable time evolution without repeated fermionic diagonalization, while faithfully reproducing both local dynamics and statistical correlations. These results establish gauge-invariant message passing as a compact and physically grounded framework for learning and simulating Abelian lattice gauge systems.
\end{abstract}

\maketitle

\section{introduction}

\label{sec:intro}

The growing interface between machine learning (ML) and many-body physics is reshaping how complex systems are analyzed and simulated~\cite{carleo19,karniadakis2021,bedolla21,ramprasad2017,butler2018,schmidt2019,boehnlein22}. Rather than explicitly solving microscopic models, one can construct data-driven surrogates that approximate the mapping between configurations and observables, thereby enabling access to larger system sizes and longer time scales. For such approaches to be reliable and transferable, however, they must faithfully encode the structural constraints of the underlying physical problem, among which symmetry plays a central role. Incorporating symmetry not only enforces physical consistency but also significantly reduces the effective complexity of the learning task.

For systems with global symmetries, this principle has been successfully realized through symmetry-adapted descriptors~\cite{behler07,bartok10,behler16,shapeev16,bartok13,behler11,drautz19,zhang21,zhang22,zhang23,cheng23a,Fan24,Ghosh24} and, more recently, through symmetry-aware and equivariant neural network (ENN)~\cite{cohen2016,cohen2017,cohen2018,weiler2018,anderson2019,batatia2022,batzner2022,musaelian2023,gong2023,yang2025,batatia2025,Fan2026,kondor2025} architectures that explicitly enforce invariance or equivariance under the relevant transformations. These approaches restrict the hypothesis space to physically admissible functions, improving both efficiency and generalization. Graph neural networks (GNNs)~\cite{gilmer2017,hamilton2017,xie2018,schutt2018,xu2019,maron2019,Chen2019,unke2019,choudhary2021,reiser2022,dai2021,Fan2026b} are particularly well suited to this setting: by organizing degrees of freedom on vertices and interactions on edges, they naturally encode locality, weight sharing, and permutation symmetry. This combination has led to successful applications across molecular modeling, materials science, and lattice many-body systems.

Local gauge symmetry presents a fundamentally different situation~\cite{wilson1974,kogut1975,kogut1979,creutz1983,rothe2012}. Unlike global symmetries, it reflects a redundancy in the description: multiple configurations correspond to the same physical state through independent transformations at each lattice site. As a result, the physically relevant information is not contained in the elementary variables themselves, but in gauge-invariant combinations. Any learning framework that operates directly on gauge-dependent variables must therefore either learn to ignore this redundancy or enforce it explicitly, both of which can introduce inefficiency. A more natural strategy is to construct representations that live entirely within the gauge-invariant sector from the outset.

Such considerations are central to Abelian lattice gauge theories, where link variables take values in groups such as $\mathbb{Z}_2$ or $\mathrm{U}(1)$ and observables are built from closed-loop operators and fluxes. While originally developed in high-energy physics, these structures arise broadly in condensed matter~\cite{wen1991,wen2002,hermele2004,lee2006,savary2017,zhou2017,wen2004,Fradkin2013}. Emergent $\mathbb{Z}_2$ gauge fields appear in quantum spin liquids, most notably in the Kitaev honeycomb model, where fractionalized excitations couple to static gauge backgrounds~\cite{Kitaev2006,Hermanns2018}. Related gauge descriptions also arise in quantum dimer models and frustrated magnets, while spin-ice systems exhibit an emergent $\mathrm{U}(1)$ gauge structure with Coulomb-phase correlations. In parallel, advances in atomic, molecular, and optical (AMO) platforms have enabled direct realizations of lattice gauge models through quantum link constructions~\cite{chandrasekharan1997,brower1999}, where gauge fields are represented by finite-dimensional operators and their dynamics can be experimentally probed~\cite{wiese2013,wiese2022,banerjee2012,zohar2013,tagliacozzo2013,zohar2016,zhou2022,zohar2013b,cheng2024,surace2020}. Across these settings, extracting physical information requires working with gauge-invariant observables.

Existing ML approaches to lattice gauge systems have largely emphasized architectures that maintain covariance under local gauge transformations~\cite{favoni2022,bulusu2021,lehner2023,apte2024,cohen2019,boyda2021,albergo2019,kanwar2020,nagai2025,luo2021,Spriggs2026,Gert2026}. While formally appealing—and in fact likely essential for non-Abelian theories, where the noncommutative structure and richer loop algebra make a direct invariant parametrization impractical—such constructions operate on redundant variables and require explicit tracking of transformation properties throughout the network. By contrast, for Abelian theories this level of structure is not strictly necessary: the space of gauge-invariant observables admits a direct and complete parametrization in terms of local loop variables and related quantities. This distinction motivates a simpler and potentially more efficient strategy for Abelian systems, in which symmetry is enforced directly at the level of representation rather than through equivariant transformations.

In this work, we develop a graph neural network framework tailored to Abelian lattice gauge theories that operates entirely within the gauge-invariant sector. The construction is based on local gauge-invariant features—most notably Wilson loops on elementary plaquettes—augmented by a minimal set of nonlocal observables such as system-spanning winding Wilson loops that capture global topological information. These features are processed by a message-passing architecture designed to respect the underlying lattice symmetries, with graph permutation symmetry enforcing discrete lattice translations and point-group operations. Because all inputs are gauge invariant by construction, the model eliminates redundancy associated with gauge degrees of freedom while retaining both expressive power and favorable scalability.

We benchmark the proposed approach on both $\mathbb{Z}_2$ and $\mathrm{U}(1)$ lattice gauge models, demonstrating accurate prediction of local observables and global quantities across a range of configurations. Beyond static tasks, we further employ the learned model as an efficient surrogate for semiclassical gauge-field dynamics, enabling large-scale simulations relevant to quantum link models and synthetic gauge systems in AMO platforms. These results establish gauge-invariant message passing as a compact and physically grounded framework for learning and simulating Abelian lattice gauge systems.

The rest of the paper is organized as follows. In Sec.~II, we introduce the architecture of the gauge-invariant GNN, including the construction of node features and symmetry-preserving message-passing rules. In Sec.~III, we benchmark the model on a gauge–matter tight-binding system, demonstrating accurate predictions of both total energy and local fermion density. In Sec.~IV, we assess dynamical performance in a $\mathrm{U}(1)$ quantum link model (gauge magnet), where the learned model serves as a surrogate for semiclassical dynamics. Finally, Sec.~V summarizes the results and provides an outlook.

\section{Gauge-invariant Graph Neural networks }

\label{sec:GNN}

To make the discussion concrete, we focus on Abelian lattice gauge theories and develop a machine-learning framework that operates entirely within the gauge-invariant sector. While gauge-equivariant constructions provide a general route for incorporating local symmetry, Abelian systems admit a more direct formulation in terms of gauge-invariant variables. This enables one to eliminate redundant gauge degrees of freedom at the outset and instead use neural networks to capture spatial correlations and lattice symmetries encoded in physical observables. Our goal is therefore to construct a graph neural network (GNN) based on gauge-invariant representations, with particular emphasis on settings where gauge fields mediate nonlocal correlations.

We consider Abelian lattice gauge theories with gauge group $\mathbb{Z}_N$ (including the paradigmatic $\mathbb{Z}_2$ case) or $\mathrm{U}(1)$. The gauge degrees of freedom reside on oriented links $(ij)$ and are parametrized by phase variables $e^{i\theta_{ij}}$, with the convention $\theta_{ji} = -\theta_{ij}$. A local gauge transformation is specified by site-dependent phases $\alpha_i$, under which the link variables transform as
\begin{equation}
	e^{i\theta_{ij}} \;\longrightarrow\; e^{i\theta'_{ij}} = e^{i(\theta_{ij} + \alpha_i - \alpha_j)}.
\end{equation}
For $\mathbb{Z}_N$, the phases $\theta_{ij}$ and $\alpha_i$ take discrete values in multiples of $2\pi/N$. These transformations reflect the local redundancy of the gauge description, and physical observables must therefore be invariant under them.

In Abelian theories, all gauge-invariant information can be expressed in terms of closed-loop variables~\cite{wilson1974,Fradkin2013}. The fundamental building blocks are Wilson loops defined on elementary plaquettes,
\begin{equation}
	\label{eq:W_square}
	W_{\square} = \exp\!\left(i \sum_{(ij)\in \square} \theta_{ij}\right),
\end{equation}
which measure the gauge flux threading each plaquette. Under a gauge transformation, the phase shifts $\alpha_i - \alpha_j$ cancel pairwise along any closed loop, ensuring that $W_{\square}$ is gauge invariant. Owing to the commutativity of the Abelian gauge group, Wilson loops associated with larger contractible contours can be systematically reduced to products of elementary plaquette loops, with contributions from internal links canceling pairwise. As a result, the set of plaquette Wilson loops $\{W_{\square}\}$ provides a complete description of all local gauge-invariant degrees of freedom.

\begin{figure*}[t]
\centering
\includegraphics[width=1.99\columnwidth]{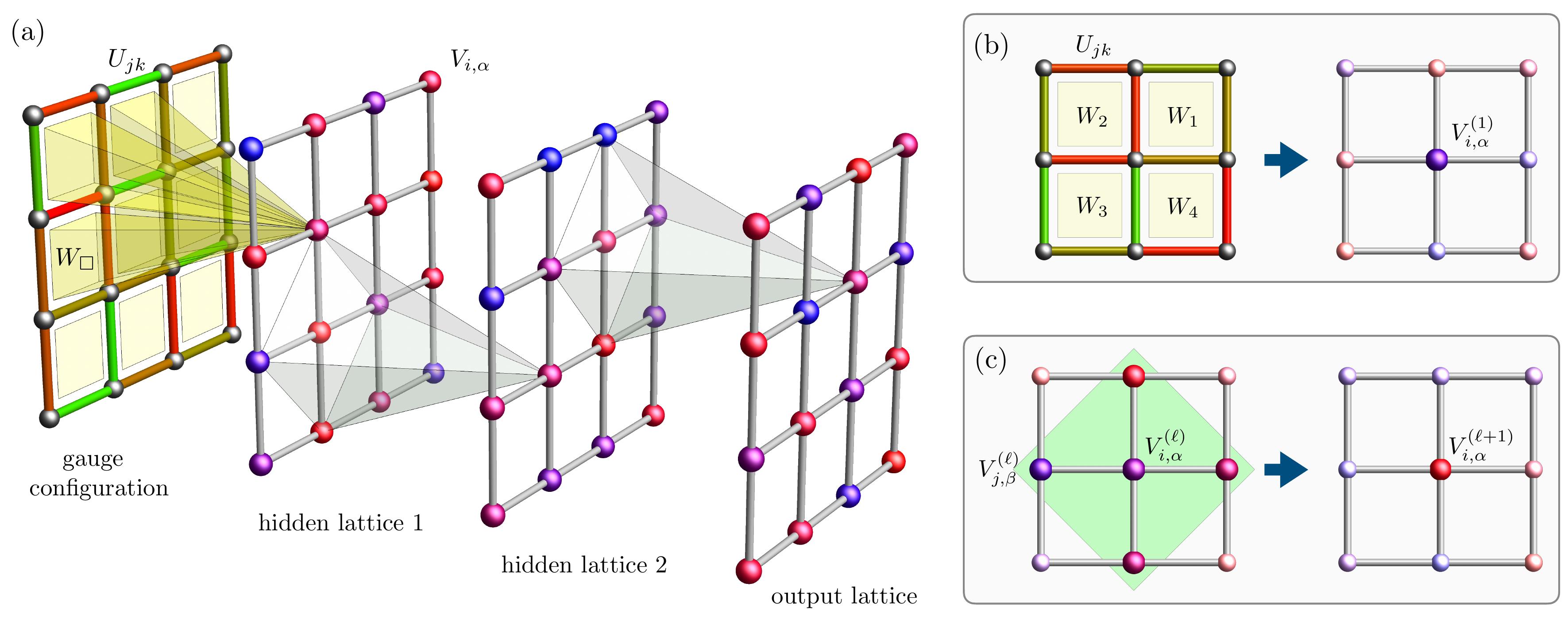}
\caption{Gauge-invariant GNN architecture.
(a) The input gauge configuration is first mapped to gauge-invariant variables given by the Wilson loops $W_\Box$ on elementary plaquettes, which are used to initialize vertex features $V_{i,\alpha}$ on the lattice graph. (b) The initial node features $V_{i, \alpha}^{(1)}$ are constructed from the local plaquette Wilson loops surrounding each site. (c) Features are then propagated across layers via standard nearest-neighbor aggregation, updating $V_{i, \alpha}^{(\ell)} \rightarrow V_{i, \alpha}^{(\ell+1)}$ while preserving permutation equivariance of the lattice graph. In this construction, gauge invariance is enforced through the choice of Wilson-loop inputs, and discrete lattice symmetries are incorporated through the graph-based message-passing structure.}
    \label{fig:GNN-scheme1}
\end{figure*}

On lattices with periodic boundary conditions, additional global degrees of freedom arise from noncontractible loops that wind around the system. These winding Wilson loops encode topological flux sectors and are independent of local plaquette variables. Together, the plaquette and winding Wilson loops furnish a complete and nonredundant parametrization of gauge-inequivalent configurations in Abelian lattice gauge theories.

In many physically relevant situations, gauge fields are coupled to additional degrees of freedom, leading to observables that depend on extended spatial correlations mediated by the gauge background. In such cases, the influence of the gauge field cannot, in general, be captured by purely local functions of a small number of Wilson loops. Instead, physical quantities reflect collective effects arising from the arrangement of fluxes across the lattice, making this setting a stringent test for any learning framework. In Sec.~\ref{sec:struc-property}, we will introduce an explicit gauge--matter Hamiltonian to study this problem in a concrete setting.

Building on this structure, we construct a GNN that operates directly on gauge-invariant variables. The input features consist of local plaquette Wilson loops $\{W_\square\}$, optionally augmented by a small set of system-spanning winding loops to encode global information. These features are associated with plaquettes (or, equivalently, sites of the dual lattice) and are processed through a message-passing architecture defined on the lattice graph. The role of the GNN is not to enforce gauge symmetry---which is already built into the representation---but rather to incorporate spatial symmetries of the lattice. In particular, graph permutation symmetry ensures equivariance under discrete lattice translations, while weight sharing and connectivity encode point-group symmetries such as rotations and reflections.

From a computational perspective, the learning problem can be viewed as constructing a mapping from the set of local gauge-invariant variables, ${W_{\square}}$, to a target observable $\mathcal{O}[\{W_{\square} \}]$, where $\mathcal{O}$ may represent quantities such as the total action, energy, or observables arising from gauge–matter coupling. Formulating the problem in this way has two important consequences. First, by working directly with gauge-invariant degrees of freedom, the model avoids the redundancy associated with local gauge transformations and focuses entirely on physically meaningful information. This not only simplifies the representation but also improves data efficiency, since the network does not need to learn symmetry constraints implicitly. Second, when combined with a graph neural network architecture that respects the underlying lattice symmetries, the approach naturally captures spatial correlations while maintaining scalability and transferability across different system sizes and geometries.

This structure naturally leads to a fully gauge-invariant neural-network formulation. Rather than using raw link variables $e^{i\theta_{ij}}$, we construct the input directly from gauge-invariant quantities—namely, Wilson loops on all elementary plaquettes together with a minimal set of system-spanning winding loops. Since these variables provide a complete and nonredundant characterization of gauge-inequivalent configurations in Abelian lattice gauge theories, the network operates entirely within the physical (gauge-invariant) sector, and no explicit gauge-equivariance constraint is required at the architectural level.

The remaining symmetries are purely spatial and can be systematically incorporated through a graph neural network defined on the lattice. In this representation, lattice sites (or equivalently plaquettes on the dual lattice) form the nodes of a graph with nearest-neighbor connectivity. Discrete lattice translations and point-group operations act as permutations of these nodes. By constructing the GNN to be permutation-equivariant, the architecture enforces translation symmetry and respects point-group symmetries such as rotations and reflections. Gauge invariance, in contrast, is not imposed at the level of the network operations, but is guaranteed by construction through the choice of gauge-invariant input variables.

A schematic illustration of the gauge-invariant GNN is shown in Fig.~\ref{fig:GNN-scheme1}(a). The inputs consist of plaquette Wilson loops $W_\square$, from which node features are initialized. As illustrated in Fig.~\ref{fig:GNN-scheme1}(b), the feature vector at site $i$ is constructed by collecting the Wilson loops associated with the four plaquettes adjacent to that site, $\{V^{(1)}_{i,\alpha}\} = \{W_1, W_2, W_3, W_4\}$. This construction provides a local, gauge-invariant description of the surrounding gauge configuration while preserving the spatial organization of the lattice. In addition to these local features, global winding loops $W_x$ and $W_y$, which encode the topological sector of the configuration, are incorporated through global tokens (virtual nodes) that communicate with all lattice sites during message passing. This allows the network to capture both local correlations and global topological information within a unified framework.

Feature propagation proceeds via a standard message-passing scheme, as illustrated schematically in Fig.~\ref{fig:GNN-scheme1}(c). At each layer $\ell$, node features are updated by combining on-site information with contributions from nearest neighbors,
\begin{equation}
	\mathcal{X}^{(\ell)}_{i,\alpha}
	= \sum_\beta W_{\alpha\beta} V^{(\ell)}_{i,\beta}
	+ \sum_{j \in \mathcal{N}(i)} \sum_\beta W'_{\alpha\beta} V^{(\ell)}_{j,\beta},
\end{equation}
followed by a nonlinear activation,
\begin{equation}
	V^{(\ell+1)}_{i,\alpha}
	= \sigma\!\left( \mathcal{X}^{(\ell)}_{i,\alpha} + b_\alpha \right).
\end{equation}
Here, $W$, $W'$, and $b$ are learnable parameters shared across all lattice sites. This weight sharing enforces translation equivariance, while the symmetric treatment of all nearest neighbors ensures invariance under point-group operations. Repeated application of this update rule enables the network to iteratively refine node representations by incorporating information from progressively larger neighborhoods.

At the output stage, the learned node features serve as symmetry-adapted representations from which physical observables are extracted. For local observables, the multi-channel feature vector at each site is mapped to a scalar prediction via a site-wise multilayer perceptron (MLP),
\begin{equation}
	\mathcal{O}_i = \mathrm{MLP}\bigl(V^{\rm (out)}_{i,\alpha}; \bm\theta\bigr),
\end{equation}
which mixes feature channels while preserving locality and translation symmetry. Global observables can be obtained by appropriate aggregation over sites. Because the entire pipeline operates on gauge-invariant inputs, all intermediate representations and outputs remain gauge invariant by construction.

Finally, the architecture enforces a controlled notion of locality. Each message-passing layer propagates information only between neighboring sites, so the receptive field grows systematically with network depth. This provides a natural mechanism to capture increasingly long-range correlations, which is particularly important in gauge--matter coupled systems where effective interactions can extend over multiple lattice spacings. At the same time, all learnable parameters define local, site-independent mappings, ensuring both computational scalability and transferability across lattice sizes.

\section{Structure--property mapping of coupled gauge--matter systems}

\label{sec:struc-property}

\begin{table*}[t]
\centering
\setlength{\tabcolsep}{8pt}
\begin{tabular}{l | c | c || c c}
\hline\hline
\textbf{LGT model} & \textbf{Message passing (GCN) layers} & \textbf{MLP readout} & \textbf{Learning rate} & \textbf{Loss} \\
\hline

$\mathbb{Z}_2\;(\epsilon)$ 
& $512\!\times\!512\!\times\!256\!\times\!128$ 
& $128\!\times\!32\!\times\!16\!\times\!8\!\times\!1$
& $2\times10^{-4}$ & $7.898\times10^{-3}$ \\

$\mathbb{Z}_2\;(n_i)$ 
& $512\!\times\!256\!\times\!256\!\times\!128\!\times\!128$ 
& $128\!\times\!32\!\times\!16\!\times\!8\!\times\!1$
& $3\times10^{-4}$ & $1.710\times10^{-5}$ \\
\hline

U(1)$\;(\epsilon)$ 
& $512\!\times\!512\!\times\!256$ 
& $128\!\times\!64\!\times\!32\!\times\!16\!\times\!8\!\times\!1$
& $2\times10^{-4}$ & $1.009\times10^{-2}$ \\

U(1)$\;(n_i)$ 
& $512\!\times\!512\!\times\!256\!\times\!256\!\times\!128\!\times\!128$ 
& $128\!\times\!64\!\times\!16\!\times\!8\!\times\!1$
& $8\times10^{-5}$ & $1.158\times10^{-5}$ \\
\hline\hline
\end{tabular}
\caption{Network architectures and training parameters for Abelian lattice gauge theory (LGT) benchmarks. 
Message passing (GCN) layers denote graph convolutional layers defined on the lattice, and MLP readout denotes the final multilayer perceptron. 
The targets are the energy density $\epsilon = E/N$ and the local fermion density $n_i = \langle c_i^\dagger c_i \rangle$. 
Unless otherwise noted, all results are obtained after $100$ training epochs; the $\mathbb{Z}_2$ $(n_i)$ model is trained for $150$ epochs. 
Reported losses correspond to the final mean-squared error (MSE) on the test set.}
\label{tab:GNN-detail}
\end{table*}

We now apply the gauge-invariant GNN framework introduced in the previous section to concrete Abelian lattice gauge theory (LGT) models. While the formalism is general, we focus here on a minimal gauge--matter system as a representative and nontrivial benchmark. In this setting, fermions propagate in a background gauge field, and integrating out the fermionic degrees of freedom induces effective interactions among the gauge variables. These interactions are inherently nonlocal, reflecting the extended nature of fermionic wave functions, and therefore provide a stringent test for any learning framework.

The model is defined by a tight-binding Hamiltonian on a square lattice,
\begin{equation}
    H = -\tau \sum_{\langle ij \rangle} c_i^\dagger e^{i\theta_{ij}} c_j + \mathrm{h.c.},
\end{equation}
where the link variables $\theta_{ij}$ encode the gauge field. In the U(1) case, these variables can be viewed as XY spins (phases on the unit circle), while in the $\mathbb{Z}_N$ case they reduce to discrete clock variables. Although the Hamiltonian is expressed in terms of link variables, all physical observables depend only on gauge-invariant combinations, which are fully characterized by Wilson loops. Accordingly, the learning problem is formulated entirely in terms of gauge-invariant inputs $\{W_\square\}$, ensuring that the model operates directly in the physical configuration space without redundancy.

An important consequence of this formulation is that, once expressed in terms of Wilson loops, the problem reduces to a lattice system with well-defined spatial symmetries. As discussed in the previous section, the GNN architecture naturally respects lattice point-group symmetries (translations, rotations, and reflections) through weight sharing and local message passing. Combined with gauge-invariant inputs, this ensures that both gauge redundancy and lattice symmetries are incorporated at the representation level, allowing the model to focus on physically meaningful correlations.

We consider both $\mathbb{Z}_2$ and U(1) gauge fields on a $20\times20$ lattice at fixed fermion filling. The dataset is generated by sampling gauge configurations from randomized ensembles and performing exact diagonalization (ED) of the fermionic Hamiltonian for each configuration. This yields a broad set of Wilson-loop patterns and their associated fermionic responses, suitable for learning the structural–property mapping.

A key aspect of the learning problem is that the target observables fall into two distinct classes: global quantities defined over the entire lattice, and local quantities defined at each site. A prototypical global observable is the total energy,
\begin{equation}
    \label{eq:energy_E}
    E = \langle \mathcal{H} \rangle = \sum_m f(\epsilon_m)\,\epsilon_m,
\end{equation}
where $\epsilon_m$ are the single-particle eigenenergies of the tight-binding gauge--matter Hamiltonian, and $f(\epsilon)=1/(e^{\beta(\epsilon-\mu)}+1)$ is the Fermi--Dirac distribution at inverse temperature $\beta$ and chemical potential $\mu$. A representative local observable is the fermion density,
\begin{equation}
    n_i = \langle c_i^\dagger c_i \rangle = \sum_m f(\epsilon_m)\,|U^{(m)}_i|^2,
\end{equation}
where $U^{(m)}$ denotes the eigenvector corresponding to $\epsilon_m$, and $U^{(m)}_i$ is its amplitude on site $i$. Although both observables are determined by the same underlying spectrum, they impose qualitatively different requirements: global quantities involve coherent aggregation over all eigenmodes, whereas local observables require resolving spatial structure encoded in the eigenvectors, reflecting nonlocal correlations mediated by the gauge field.

These distinctions are directly reflected in the GNN design. For local observables such as $n_i$, the network produces node-wise outputs $\hat{n}_i$, trained using a site-resolved mean-squared-error loss,
\begin{equation}
    \mathcal{L}_{\mathrm{local}} = \sum_i \bigl| n_i - \hat{n}_i \bigr|^2.
\end{equation}
In this setting, the message-passing layers encode nonlocal gauge-induced correlations into local embeddings, enabling each node to approximate its corresponding quantum expectation value.

For global observables such as the total energy, we instead adopt a decomposition inspired by the Behler--Parrinello framework. The network produces site-resolved latent contributions $\hat{\varepsilon}_i$, which are summed to obtain $\hat{E} = \sum_i \hat{\varepsilon}_i$, and trained using
\begin{equation}
    \mathcal{L}_{\mathrm{global}} = \bigl| E - \sum_i \hat{\varepsilon}_i \bigr|^2.
\end{equation}
From a physical perspective, this construction can be viewed as learning an effective energy functional $E[\{W_\ell\}]$, in which the fermionic degrees of freedom have been integrated out and the energy is expressed entirely in terms of gauge-invariant loop variables.

To make this explicit, we generalize Eq.~(\ref{eq:W_square}) to arbitrary closed loops $\ell$,
\begin{equation}
    W_\ell = \exp\!\left(i \sum_{(ij)\in \ell} \theta_{ij}\right),
\end{equation}
which form a complete set of gauge-invariant variables. In Abelian gauge theories, these loop variables are not independent: any $W_\ell$ can be expressed as a product of elementary plaquette variables $W_{\square}$ over the area enclosed by the loop,
\begin{equation}
    W_\ell = \prod_{\square \in A(\ell)} W_{\square},
\end{equation}
where $A(\ell)$ denotes a surface bounded by $\ell$. This reflects the absence of nontrivial path ordering and further emphasizes that the gauge-invariant content is fully captured by local Wilson loops.

Since the energy is real and invariant under $\theta_{ij} \to -\theta_{ij}$, it depends only on $\mathrm{Re}[W_\ell]$. The effective energy can therefore be expressed as a symmetry-allowed cluster expansion,
\begin{equation}
\begin{aligned}
    E[\{W_\ell\}] 
    &= J_{1} \sum_{\square} \mathrm{Re}[W_{\square}] \\
    &\quad + J_{2} \sum_{\recttwobyone} \mathrm{Re}[W_{\recttwobyone}]
          + J_{2} \sum_{\rectonebytwo} \mathrm{Re}[W_{\rectonebytwo}] \\
    &\quad + J_{3} \sum_{\Lshape} \mathrm{Re}[W_{\Lshape}]
          + \cdots .
\end{aligned}
\end{equation}
This expression illustrates a cluster expansion of the effective energy in terms of gauge-invariant loop variables, organized by increasing spatial extent and consistent with lattice symmetries. The coefficients $J_{\mathcal{C}}$ encode fermion-mediated interactions and, in general, decay slowly with cluster size due to the extended nature of fermionic correlations. In practice, enumerating all such contributions quickly becomes intractable; the GNN can be viewed as learning this expansion implicitly, capturing contributions from loop clusters of different sizes and geometries.

\begin{figure}[t]
\centering
\includegraphics[width=0.99\columnwidth]{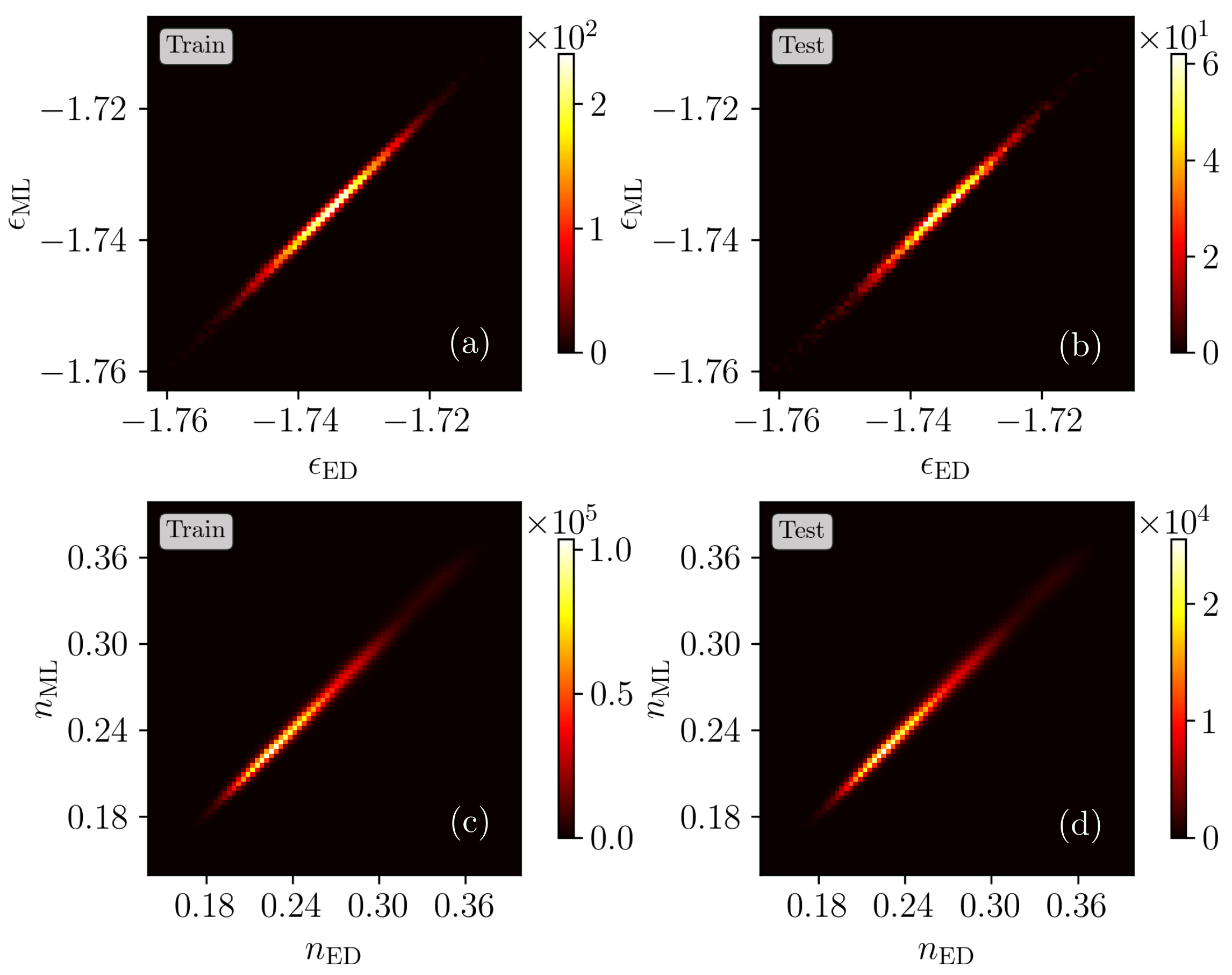}
\caption{Benchmark performance of the gauge-invariant GNN on a $20\times20$ lattice for the $\mathbb{Z}_2$ gauge model. Panels (a,b) show the energy density at half filling, and (c,d) the local fermion density at quarter filling. Heat maps compare ML predictions with exact diagonalization, with color indicating sample density; (a,c) correspond to training data and (b,d) to test data. The test accuracy reaches MSE $5.207\times10^{-7}$ and $R^2=0.9912$ for energy density, and MSE $2.245\times10^{-5}$ with $R^2=0.9830$ for fermion density.}
\label{fig:benchmark-Z2}
\end{figure}

\begin{figure}[t]
\centering
\includegraphics[width=0.99\columnwidth]{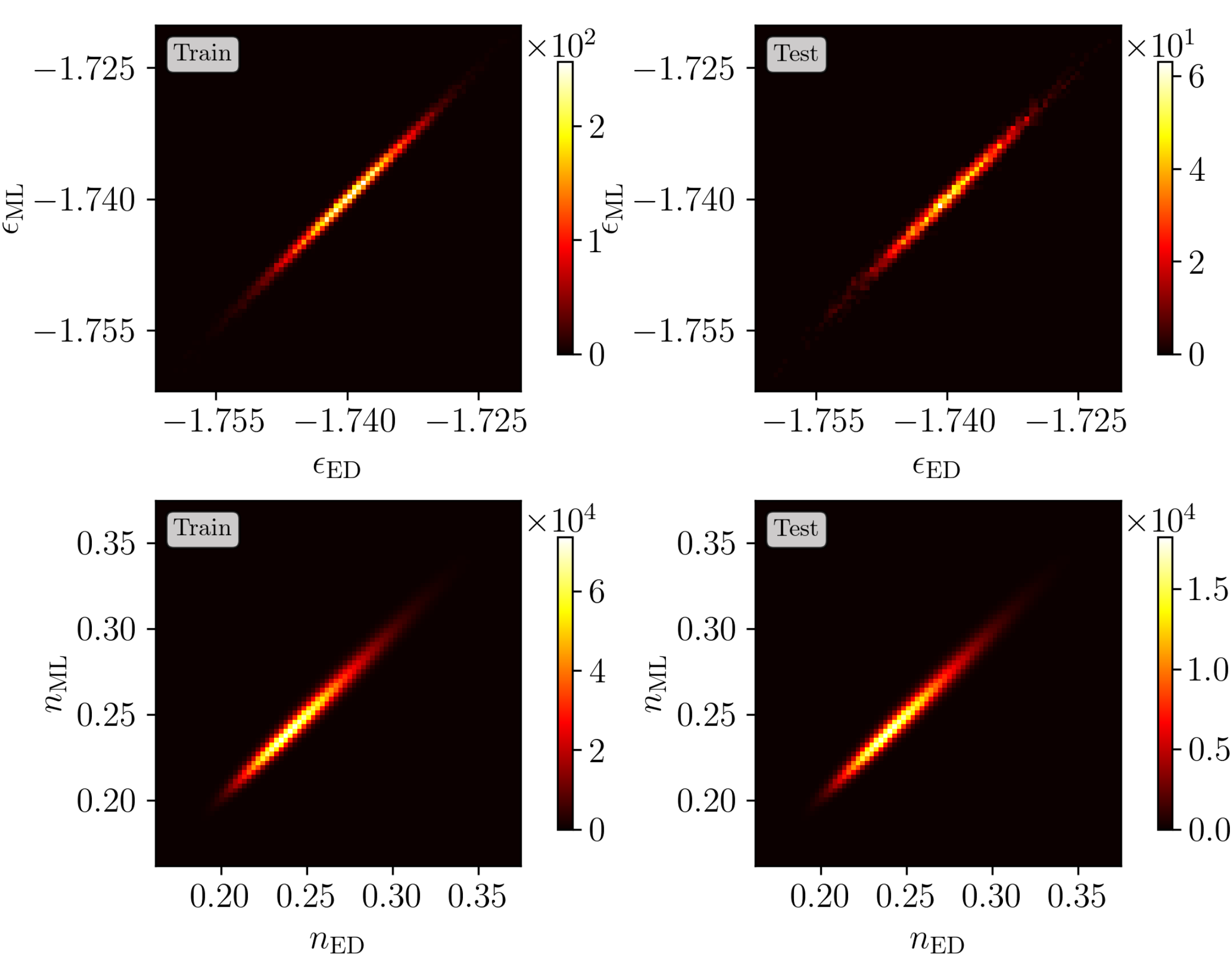}
\caption{Performance of the gauge-invariant GNN on a $20\times20$ lattice for the U(1) gauge model. Panels (a,b) display the energy density at half filling, while (c,d) show the local fermion density at quarter filling. Results are presented as density heat maps comparing ML predictions to exact diagonalization; the left column (a,c) shows training data and the right column (b,d) test data. The test performance yields MSE $4.6165\times10^{-7}$ and $R^2=0.9826$ for energy density, and MSE $1.572\times10^{-5}$ with $R^2=0.9747$ for fermion density.}
\label{fig:benchmark-U1}
\end{figure}

The network architecture and training parameters are summarized in Table~\ref{tab:GNN-detail}. The predictive performance is shown in Figs.~\ref{fig:benchmark-Z2} and \ref{fig:benchmark-U1}, where parity plots compare GNN predictions with exact diagonalization for both energy density and local fermion density. In all cases, the data collapse tightly along the diagonal, with minimal scatter and no visible systematic bias. The $\mathbb{Z}_2$ and U(1) models achieve similarly high accuracy across both global and local observables.

The accuracy obtained for the local fermion density is particularly notable. Unlike the total energy, which is a single scalar, $n_i$ encodes spatially resolved quantum expectation values that depend sensitively on nonlocal interference effects associated with fermion propagation through extended Wilson-loop configurations. The ability of the GNN to reproduce these quantities demonstrates that it captures nontrivial gauge-invariant correlations beyond purely local features.

Overall, this benchmark shows that the gauge-invariant GNN provides an accurate and scalable surrogate for structure--property mappings in coupled gauge--matter systems. By combining gauge-invariant representations with symmetry-aware message passing, the model captures both local structure and long-range correlations, offering a physically grounded and flexible framework for learning in lattice gauge theories.

\section{Semiclassical dynamics of gauge magnets}

Having established the accuracy of the gauge-invariant GNN for static structure--property mappings, we now turn to a more demanding dynamical benchmark. In the previous section, the network was trained to predict observables such as the total energy and local fermion density from gauge-invariant Wilson-loop configurations. Here, we assess whether the same framework can be extended to model real-time semiclassical dynamics, where the network serves as a surrogate for the effective force field governing gauge evolution. This constitutes a significantly more stringent test: small errors in local forces can accumulate under time integration, and faithful long-time dynamics requires an accurate representation of the underlying energy landscape.

To this end, we benchmark the method using a $U(1)$ quantum link model (QLM)~\cite{zohar2013b,cheng2024,surace2020}, which provides a minimal and physically transparent setting for dynamical lattice gauge theories. QLMs offer a finite-dimensional formulation in which the gauge degrees of freedom on each link are represented by quantum spins, rather than continuous rotor variables. In the $U(1)$ case, the electric field is identified with the $S^z$ component of the spin, while the gauge connection is encoded in the raising and lowering operators $S^\pm$. This construction preserves local gauge invariance while replacing the infinite-dimensional Hilbert space of conventional lattice gauge theory with a finite one. In this representation, the dynamical gauge fields behave as interacting quantum spins residing on the links of the lattice, and the model is therefore often referred to as a \emph{gauge magnet}~\cite{wiese2022,wiese2013,zohar2016}. The terminology reflects the fact that the gauge sector can be viewed as a generalized spin system with local constraints (Gauss law) and ring-exchange interactions generated by plaquette terms, a perspective that is particularly useful for semiclassical dynamics.

We consider an idealized $U(1)$ quantum link model that retains the essential ingredients of matter--gauge coupling, electric-field energy, and magnetic (plaquette) dynamics. The gauge field on each oriented link $\langle ij\rangle$ is represented by a spin variable, and the Hamiltonian is
\begin{equation}
\begin{aligned}
	\mathcal{H} =&-\tau \sum_{\langle ij\rangle} \left(c_i^\dagger S^+_{ij} c_j+\mathrm{h.c.}\right)	
	+\frac{g}{2}\sum_{\langle ij\rangle}(S^z_{ij})^2 \\
	&-\frac{K}{2}\sum_{\square}	\left(W_{\square}+W_{\square}^\dagger\right),
\end{aligned}
\end{equation}
where $c_i^\dagger$ creates a fermion on site $i$. The first term describes gauge-invariant hopping of fermions, with $\tau$ setting the kinetic energy scale. The second term represents the electric-field energy, with coupling $g$ controlling the cost of nonzero electric flux $S^z_{ij}$. The third term encodes magnetic (flux) dynamics with strength $K$, where
\begin{equation}
    W_{\square}=S^+_{ij}S^+_{jk}S^-_{lk}S^-_{il}.
\end{equation}
The link operators obey the standard spin algebra,
\begin{equation}
	[S^z_{ij},S^\pm_{ij}]=\pm S^\pm_{ij},
\qquad
	[S^+_{ij},S^-_{ij}]=2S^z_{ij}.
\end{equation}

Physical states must satisfy the Gauss law constraint,
\begin{equation}
	\sum_{j\in\mathcal{N}(i)} \eta_{ij} S^z_{ij} = n_i - \rho_i^{\mathrm{bg}},
\end{equation}
where $\eta_{ij}$ specifies the orientation of the link relative to site $i$. A convenient choice of initial conditions is to set $S^z_{ij}=0$ and fix the fermion density such that $n_i = \rho_i^{\mathrm{bg}}$. The model is invariant under local $U(1)$ gauge transformations,
\begin{equation}
	c_i \rightarrow e^{i\alpha_i}c_i,
\qquad
	S^+_{ij} \rightarrow e^{i(\alpha_i-\alpha_j)} S^+_{ij}.
\end{equation}

In the semiclassical limit, each link spin is treated as a classical vector of fixed length, parameterized by polar coordinates $(\theta_{ij}, \phi_{ij})$. In this representation, the spin components are expressed as
\begin{equation}
    S^\pm_{ij} = S \sin\theta_{ij}\, e^{\pm i\phi_{ij}}, 
    \qquad
    S^z_{ij} = S \cos\theta_{ij}.
\end{equation}
This classical description can be viewed as the large-$S$ limit of the underlying quantum dynamics. In this limit, the operator commutators listed above are replaced by Poisson brackets, which govern the classical dynamics of $\mathbf{S}_{ij}$. The resulting equations of motion take the form
\begin{equation}
	\frac{d\mathbf{S}_{ij}}{dt} = \mathbf{S}_{ij}\times \mathbf{H}_{ij}, 
\end{equation}
where
\begin{equation}
	\mathbf H_{ij} = -\frac{\partial E}{\partial \mathbf S_{ij}},
\end{equation}
and $E = \langle \mathcal{H} \rangle$ denotes the adiabatic energy obtained after integrating out the fermionic degrees of freedom, with its explicit expression in terms of ED eigenenergies given in Eq.~(\ref{eq:energy_E}). For the transverse components, it is convenient to consider derivatives with respect to $S^\pm$, $H^\pm_{ij} = -\frac{\partial E}{\partial S^\pm_{ij}}$, which yields
\begin{equation}
    \label{eq:H+} 
    H^+_{ij} = -\tau \rho_{ij} -\frac{K}{2} \Big( S^+_{jk} S^-_{kl} S^-_{il} + S^-_{mj} S^-_{mn} S^+_{ni} \Big),
\end{equation}
with $H^-_{ij} = \bigl( H^+_{ij} \bigr)^*$. Here $\rho_{ij}$ denotes the single-particle density matrix,
\begin{equation}
    \rho_{ij} = \langle c_i^\dagger c_j \rangle = \sum_m f(\epsilon_m)\, U^{(m)*}_i U^{(m)}_j,
\end{equation}
where $f(\epsilon)$ is the Fermi--Dirac distribution, and $\epsilon_m$ and $U^{(m)}$ are the fermionic eigenvalues and eigenvectors. Its appearance highlights the nonlocal character of the force field, as $\rho_{ij}$ encodes correlations mediated by extended fermionic states, implying that direct simulation requires repeated exact diagonalization at each time step. The longitudinal component takes the simpler form
\begin{equation}
    \label{eq:Hz}
    H^z_{ij} = g S^z_{ij}.
\end{equation}
which is purely local and follows directly from the explicit $S^z$ dependence of the effective energy. In contrast to the transverse components, $H^z_{ij}$ does not couple to fermionic bilinears and therefore does not inherit the nonlocality associated with $\rho_{ij}$. As a result, the longitudinal contribution acts as a local restoring term set by $g$ and can be evaluated efficiently without additional fermionic calculations. This highlights that the computational bottleneck in evaluating $ H^\pm_{ij}$ arises from the transverse sector.

The construction of a force model within the present gauge-invariant framework requires particular care. Since the GNN input consists solely of gauge-invariant Wilson loops $\{W_\square\}$, its output must also be gauge invariant. Consequently, the network cannot directly represent gauge-covariant quantities such as the local field ${H}^\pm_{ij}$ that enters the equations of motion.

To overcome this limitation, we adopt a strategy analogous to the Behler--Parrinello (BP) scheme. The GNN is trained to predict a gauge-invariant local energy density $\varepsilon_i$, from which the total energy is constructed as $E = \sum_i \varepsilon_i$. This defines an effective energy functional $E[\{W_\square\}]$ entirely in terms of gauge-invariant variables. The gauge-covariant field is then obtained via automatic differentiation with respect to the underlying link variables. In this way, although the model operates strictly within the gauge-invariant sector, the differentiation step naturally generates the required gauge-covariant information.

Following this paradigm, we train the GNN using the local field as the target observable. The loss function is defined as
\begin{equation}
	\mathcal{L} = \sum_{\langle ij\rangle} \left| H^+_{ij}  - \hat{H}^+_{ij} \right|^2,
\end{equation}
where ${H}^+_{ij}$ is obtained from ED via Eq.~(\ref{eq:H+}), and $\hat{H}^+_{ij}$ is computed from the learned energy through automatic differentiation. This formulation provides richer supervision than energy-based training alone while ensuring consistency with an underlying energy functional, which is crucial for stable long-time dynamics.

Training data were generated from semiclassical simulations on a $20 \times 20$ lattice using 50 ensembles of random initial conditions. Each trajectory was evolved for $8{,}000$ time steps with $\Delta t = 10^{-3}$ using semi-implicit and Cayley integration schemes, which ensure numerical stability and good energy conservation. Configurations were sampled every 20 steps, yielding approximately $2\times10^4$ samples, which were split into training and test sets (80/20). All simulations were performed at near-zero temperature to avoid unphysical energy drift.
The GNN architecture consists of four graph-convolution layers with channel dimensions $(256,1024,512,256)$ and GELU activations, followed by an MLP readout $(256,128,64,32,16,4)$, for a total of $1.88\times10^6$ parameters. Training was performed using AdamW with weight decay $10^{-3}$ and a cosine annealing schedule over 100 epochs. The final model achieves an MSE of $2.45\times10^{-3}$ and $R^2=0.9978$ on both training and test sets, indicating excellent generalization.

\begin{figure}[t]
\centering
\includegraphics[width=0.99\columnwidth]{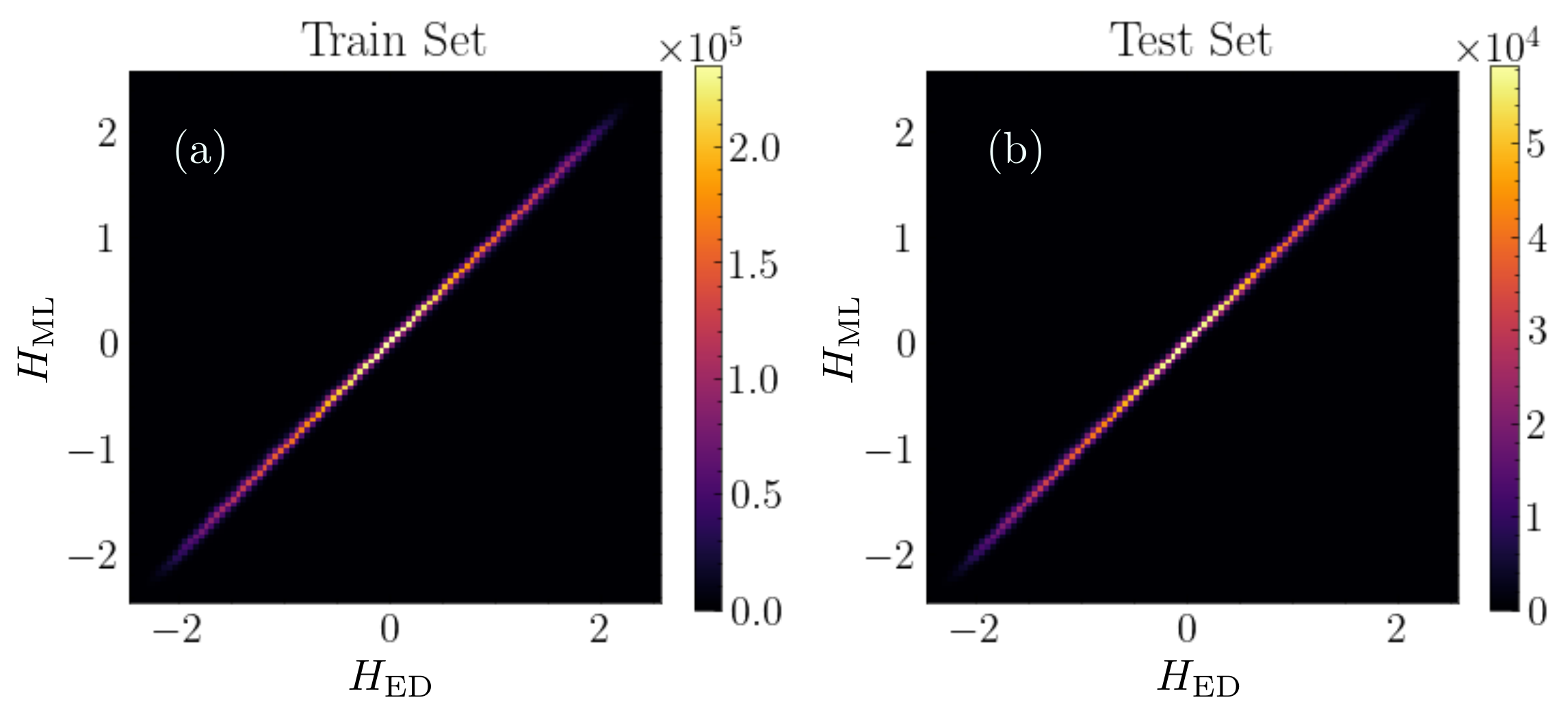}
\caption{Benchmark of the gauge-invariant GNN for local force prediction in the $\mathrm{U}(1)$ gauge magnet. Panels (a) and (b) show density heat maps comparing ML-predicted and exact forces for the training and test sets. The strong diagonal concentration indicates excellent agreement, with $R^2 \approx 0.998$, MSE $\sim 2.45\times 10^{-3}$, and MAE $\sim 3.46\times 10^{-2}$, demonstrating accurate reconstruction of the effective force field governing the semiclassical dynamics.}    \label{fig:benchmark-force}
\end{figure}

We begin by assessing the quality of the learned force field at the most immediate level, namely through instantaneous force predictions. Fig.~\ref{fig:benchmark-force} presents parity plots comparing the ML-predicted local fields with those obtained from exact diagonalization (ED) for both the training and test datasets. As seen in Fig.~\ref{fig:benchmark-force}, the data points are tightly clustered along the diagonal over the entire range of field values, with no visible systematic deviations. This indicates that the network accurately reconstructs the local effective fields across configurations sampled from the dynamical trajectories, and that the model generalizes well beyond the training set. The near-identical behavior between training and test results further confirms that overfitting is negligible and that the learned mapping captures the underlying physical structure of the force field.

\begin{figure}[t]
\centering
\includegraphics[width=0.99\columnwidth]{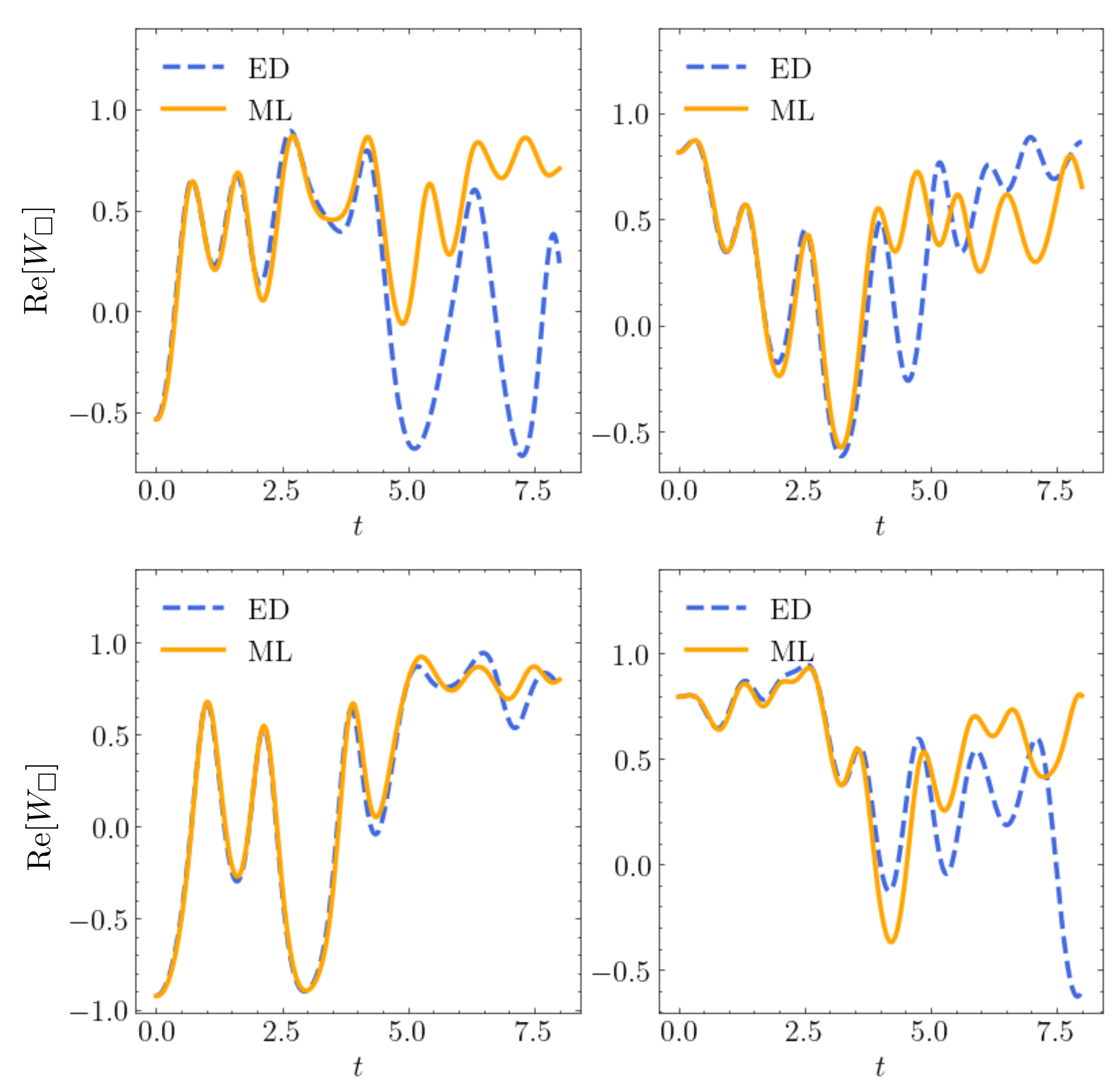}
\caption{Comparison of semiclassical dynamics in the $\mathrm{U}(1)$ quantum link model (gauge magnet). The time evolution of the real part of a Wilson loop, $\mathrm{Re}[W_{\square}(t)]$, evaluated on a randomly selected elementary plaquette, is shown for several representative trajectories. Solid lines (ML) are obtained using forces predicted by the gauge-invariant GNN, while dashed lines (ED) correspond to forces computed from exact diagonalization. The ML-driven dynamics accurately reproduce the short-time behavior and overall oscillatory structure, while small deviations accumulate at longer times, consistent with typical trajectory divergence in learned dynamical systems.}
    \label{fig:trajectories}
\end{figure}

\begin{figure}[t]
\centering
\includegraphics[width=0.99\columnwidth]{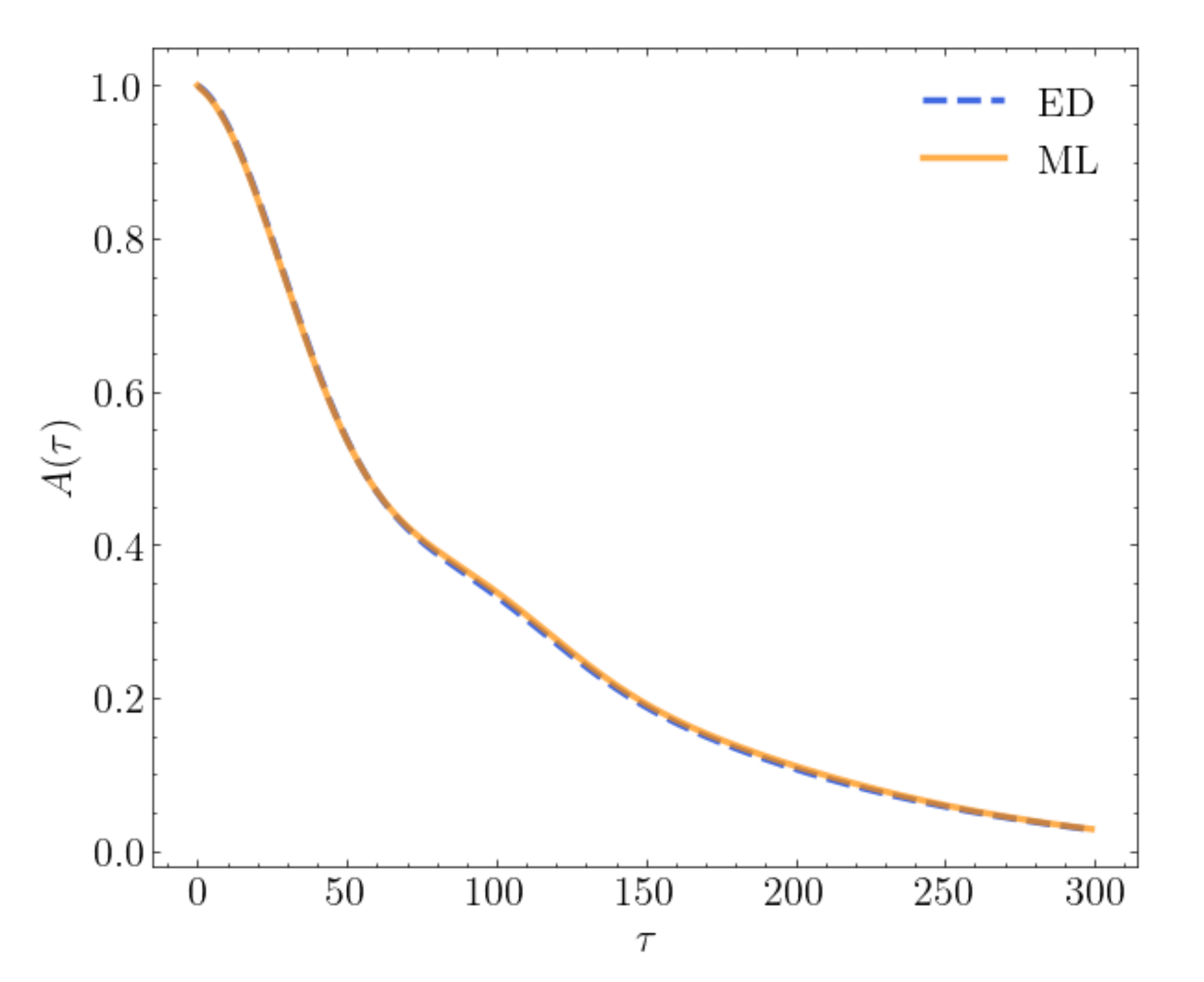}
\caption{Autocorrelation function of Wilson-loop dynamics in the $\mathrm{U}(1)$ quantum link model. The quantity $A(\tau) = \langle \mathrm{Re}[W_{\square}(t)],\mathrm{Re}[W_{\square}(t+\tau)] \rangle$ is computed by averaging over all plaquettes and multiple independent trajectories. Results obtained using forces from exact diagonalization (ED, dashed line) and from the gauge-invariant GNN (ML, solid line) are shown. The close agreement over the full time range demonstrates that, despite small trajectory-level deviations, the ML-driven dynamics accurately capture the statistical and temporal correlations of the underlying gauge-field evolution.}
    \label{fig:autocorr}
\end{figure}

Having established the accuracy of instantaneous predictions, we next evaluate the performance of the model in actual dynamical simulations. Fig.~\ref{fig:trajectories} shows representative time traces of $\mathrm{Re}[W_\square(t)]$ for several randomly selected plaquettes, comparing trajectories obtained using ED-derived forces and those generated by the ML model. At short times, the agreement is essentially indistinguishable, with the ML trajectories faithfully reproducing both the amplitude and phase of the oscillations. As time evolves, small deviations gradually emerge, reflecting the cumulative effect of small local force errors in a nonlinear dynamical system. Nevertheless, the ML dynamics continues to capture the dominant oscillatory behavior and characteristic time scales of the exact evolution, indicating that the learned force field provides a quantitatively accurate approximation to the true dynamics over physically relevant time windows.

To further assess the fidelity of the learned dynamics at a statistical level, we compute the autocorrelation function
\begin{equation}
	A(\tau)=\langle \mathrm{Re}[W_\square(t)]\,\mathrm{Re}[W_\square(t+\tau)] \rangle,
\end{equation}
averaged over all plaquettes and independent trajectories. As shown in Fig.~\ref{fig:autocorr}, the autocorrelation functions obtained from ML-driven and ED-driven simulations are nearly indistinguishable across the entire time range. In particular, both the initial decay and the long-time tail are reproduced with high accuracy, demonstrating that the model correctly captures the temporal correlations and relaxation behavior of the gauge-field dynamics. This agreement is especially notable given the trajectory-level deviations observed at long times, and highlights that the learned model preserves the statistical structure of the dynamics even when individual trajectories begin to diverge.

Overall, these results demonstrate that the gauge-invariant GNN provides an accurate and robust surrogate for semiclassical dynamics in gauge magnets. By learning an effective energy functional entirely within the gauge-invariant sector, the model enables stable and scalable time evolution without the need for repeated fermionic diagonalization at each time step, while faithfully reproducing the underlying force field. This leads to quantitatively accurate short-time trajectories and, more importantly, preserves the correct statistical structure of the dynamics at longer times. Such a combination of local accuracy and statistical fidelity is essential for practical simulations of interacting gauge systems, where exact methods are computationally prohibitive. The present results therefore establish that symmetry-constrained, energy-based GNNs can capture both the microscopic structure of effective forces and the emergent dynamical behavior, providing a viable and efficient route for large-scale simulations of lattice gauge dynamics beyond the reach of conventional approaches.

\section{Conclusion and Outlook}
\label{sec:discussion}

In this work, we have developed a gauge-invariant graph neural network (GNN) framework tailored to Abelian lattice gauge theories, in which symmetry is enforced directly at the level of representation through Wilson-loop variables. By operating entirely within the gauge-invariant sector, the model eliminates redundant degrees of freedom while retaining the ability to capture spatial correlations via symmetry-aware message passing. Benchmarks on both $\mathbb{Z}_2$ and $U(1)$ gauge–matter systems demonstrate that the GNN accurately predicts both global observables, such as the total energy, and local quantities, such as fermion densities, despite the inherently nonlocal structure induced by integrating out fermions. As discussed in Sec.~III, the network effectively learns a gauge-invariant energy functional in terms of loop variables, providing a compact and implicit realization of the underlying cluster expansion. 

We further extended this framework to semiclassical dynamics in a $U(1)$ quantum link model, where the GNN serves as a surrogate for the effective force field. By combining a gauge-invariant energy representation with automatic differentiation, the approach enables the reconstruction of gauge-covariant forces while maintaining consistency with an underlying energy functional. The resulting ML-driven dynamics accurately reproduce short-time trajectories and, more importantly, capture the statistical properties of the evolution, such as autocorrelation functions, without requiring repeated exact diagonalization. This establishes gauge-invariant GNNs as a practical and scalable approach for dynamical simulations of gauge systems.

Looking forward, the present framework opens several promising directions. A natural application is its integration with self-learning Monte Carlo (SLMC) schemes~\cite{Liu2017,Huang2017} for lattice gauge systems with fermionic matter. In such approaches, Monte Carlo updates are performed in the space of gauge configurations, but each update typically requires solving a fermionic problem (e.g., via exact diagonalization), which constitutes the dominant computational cost~\cite{abbott2022,nagai2023,nagai2024}. Within the present formulation, the GNN provides a direct surrogate for the gauge-invariant energy functional $E[\{W_\square\}]$, and can therefore be used to accelerate or even replace the expensive fermionic evaluation during sampling. This strategy is particularly appealing for systems with emergent gauge structures, such as the Kitaev honeycomb model, where Monte Carlo sampling of flux configurations is coupled to a quadratic Majorana fermion problem~\cite{Nasu2014,Eschmann2020,yang2025b}. In this context, the GNN offers a route to bypass repeated fermionic diagonalization while preserving the nonlocal dependence of observables on flux backgrounds.

Another direction concerns extensions to fully quantum lattice gauge models. While the present dynamical results are restricted to semiclassical (adiabatic) regimes, the underlying idea of learning gauge-invariant energy functionals suggests possible connections to quantum simulation frameworks. For example, one may construct variational wave functions based on GNN representations of gauge-invariant degrees of freedom, to be used within variational Monte Carlo (VMC) for ground-state studies~\cite{luo2021,Spriggs2026,luo2023,Han2021,cheng2025}. Similarly, time-dependent extensions based on the time-dependent variational principle~\cite{Dirac1930,Kramer1981,Haegeman2011} could enable approximate real-time dynamics within a reduced, symmetry-constrained manifold. Such approaches would provide an alternative route to tackling quantum gauge systems beyond semiclassical limits.

Finally, the broader implication of this work lies in the demonstration that symmetry-constrained learning at the level of physical degrees of freedom—rather than at the level of redundant variables—can lead to both conceptual and computational simplifications. While this strategy is particularly natural for Abelian gauge theories, it also provides a useful complement to gauge-equivariant constructions, and may inspire hybrid approaches in more general settings. Together, these directions point toward a unified framework in which machine learning serves not merely as a numerical tool, but as a physically structured representation of gauge-theoretic many-body systems.

\par\vspace*{6pt}

\begin{acknowledgments}
This work was supported by the U.S. Department of Energy, Office of Basic Energy Sciences, under Contract No.~DE-SC0020330. The authors thank Yaohang Li for collaboration on related work, and M.~Engelhardt for insightful discussions on lattice gauge theories. The authors also acknowledge Research Computing at the University of Virginia for providing computational resources and technical support that contributed to this work.
\end{acknowledgments}

\bibliography{ref.bib}

\end{document}